\documentstyle[twoside,fleqn,espcrc2]{article}

\newcommand{\AmS}{{\protect\the\textfont2
  A\kern-.1667em\lower.5ex\hbox{M}\kern-.125emS}}

\title{Gauge model at finite temperature with massive quarks and at finite 
       density on anisotropic lattice}

\author{L.~A.~Averchenkova, V.~K.~Petrov, G.~M.~Zinovjev
        \address{Bogolyubov Institute for Theoretical Physics,\\
        National Academy of Sciences of Ukraine, Kiev 143, UKRAINE}}
       
\begin{document}

\begin{abstract}
Critical properties of QCD and the chiral condensate at finite density are
analytically studied on an anisotropic lattice in the approximation
$SU(N)\simeq Z(N)$. Asymptotic behavior of the partition function and 
its continuum limit are discussed. 
\end{abstract}

\maketitle

In the previous paper \cite{averch} we considered the pure gluon sector of
QCD at finite temperature on lattice with the anisotropy 
$\xi \equiv a_{\sigma}/a_{\tau}\neq 1$ [where $a_{\sigma} (a_{\tau})$ is
the spatial (temporal) spacing]. At present the use of anisotropic lattices
$\xi >>1$ becomes a very popular technique \cite{anis}. In this paper 
we include matter fields into the consideration. The approximation 
$SU(N)\simeq Z(N)$ used here does not cover all features of the SU(N) 
gauge theory, but since the SU(N) and Z(N) theories belong to the same 
universality class \cite{S-Y}, it is commonly believed that the Z(N) 
degrees of freedom are responsible for many important aspects of the 
SU(N) phase structure \cite{SU-Z}.

\section{Z(N) model on anisotropic lattice}

Here we consider lattice QCD in the Hamiltonian limit where all the terms 
proportional to $1/\xi$ are neglected \cite{averch,P}, so the partition 
function can be written as
\begin{equation}
{\cal Z}=\sum_{\{\sigma \},\{\psi \},\{\bar{\psi}\}}
\exp \left\{ -{\cal S}^{G}-{\cal S}^{F} \right\}   
\label{1}
\end{equation}
with the gluon action 
\[
-{\cal S}^{G} = \kappa_{\tau} \sum_{\vec{x},\tau;n}
\sigma_{\vec{x},\tau;0}\sigma_{\vec{x},\tau +0;n}
\sigma_{\vec{x}+n,\tau;0}^{*}\sigma_{\vec{x},\tau;n}^{*} + c.c.,
\]
where $\sigma_{\vec{x},\tau;\nu} \in Z(N)$, and 
$\kappa_{\tau}\equiv \frac{2N}{g_{\tau}^2}\xi$. For the fermion part of 
the action, we choose the form \cite{HK,BRWS} in the approximation 
$\xi >> 1$
\begin{equation}
-{\cal S}^{F}=n_{f}a_{\sigma}^{3}\sum_{x}
\bar{\psi}_{x^{\prime}}D_{x^{\prime}x}^{0}\psi_{x},  
\label{2}
\end{equation}
\begin{eqnarray}
D_{x^{\prime}x}^{0} &=& 
\frac{r-\gamma_0}{2}\sigma_0(x)\delta_{x}^{x^{\prime}-0} \label{3} \\
&+& \frac{r+\gamma_0}{2}\sigma_0^*(x)\delta_{x}^{x^{\prime}+0} -
(r+ma_{\tau})\delta_{x}^{x^\prime},  
\nonumber
\end{eqnarray}
where $r$ is the Wilson parameter $(0 < r \leq 1)$, $n_{f}$ is the
number of flavors.

For gluon fields we use the periodic boundary conditions 
$\sigma_{\vec{x},\tau;\mu}=\sigma_{\vec{x},\tau +N_{\tau};\mu}$ and fix the
Hamiltonian gauge 
$\sigma_{\vec{x},\tau;0}=1+\delta_{\tau}^{0}(\Omega_{\vec{x}}-1)$;
$\Omega_{\vec{x}}=\prod_{\tau =0}^{N_{\tau}-1}\sigma_{\vec{x},\tau;0}$ 
is the Polyakov loop. Fermion fields obey
standard antiperiodic boundary conditions.

Summing over the spatial variables $\sigma_{x;n}$ \cite{averch,P}, one may
easily get 
\begin{equation}
-{\cal S}_{eff}^{G} = \tilde{\kappa}_{\tau}(N_{\tau}) \sum_{\vec{x};n}
\Omega_{\vec{x}}\Omega_{\vec{x}+n}^{*} + c.c. ,
\label{4}
\end{equation}
where the effective coupling $\tilde{\kappa}_{\tau}$ is defined as 
\[
\frac{1-e^{-3\tilde{\kappa}_{\tau}}}{1+2e^{-3\tilde{\kappa}_{\tau}}} =
\left \{\frac{1-e^{-3{\kappa}_{\tau}}}{1+2e^{-3{\kappa}_{\tau}}} \right \}^ 
{N_{\tau}}.
\]

The effective action (\ref{4}) coincides with the action
obtained in \cite{GKO} in a strong coupling approximation where the
magnetic part of the action is suppressed by factor $g^{-2}$. On an extremely
anisotropic lattice $(\xi \rightarrow \infty)$, the similar
suppression appears due to factor $\xi^{-1}$, the model may be studied
beyond the strong coupling area \cite{averch,P,billo} and the limit 
$a_{\tau}<< a_{\sigma}\rightarrow 0$ may be monitored.

In the weak coupling region, the effective coupling 
$\tilde{\kappa}_{\tau}$ 
\[
\tilde{\kappa}_{\tau} \simeq -\frac{1}{3}\ln
\frac{1 - e^{-3N_\tau\cdot e^{-3\kappa_\tau}}}
     {1 + 2e^{-3N_\tau\cdot e^{-3\kappa_\tau}}}.
\]
$\lim_{N_\tau\rightarrow \infty}\tilde{\kappa}_\tau = const$ leads to the 
reasonable renormalization condition on the bare constant 
$\kappa_\tau\sim g^{-2}\sim \ln\frac{1}{\Lambda a_\tau}$.

We consider the special case $r=1$ for the fermionic action. Taking into
account that \\
$(\delta_{x-\hat{\mu},x^{\prime}})^{\dagger} = \delta_{x+\hat{\mu},x^{\prime}}$,
the action (\ref{2}) can be rewritten as 
\[
-{\cal S}^{F}=n_{f}a_\sigma^{3} \sum_{x}\left(\bar{\psi}_{x^{\prime}}^{(+)}
\Delta_{x^{\prime}x}^{\dagger}\psi_{x}^{(+)} + 
\bar{\psi}_{x^{\prime}}^{(-)}\Delta_{x^{\prime}x}\psi_{x}^{(-)} \right) 
\]
with $\psi^{\left(\pm \right)}=\frac{1\pm \gamma_{0}}{2} \psi$ and 
\[
\Delta_{x^{\prime}x}=\delta_{\vec{x}^{\prime}}^{\vec x} \left( 
\sigma_{0}(x) \delta_{t}^{t^{\prime}-1} - (1 + a_{\tau}m)
\delta_{t}^{t^{\prime}} \right).
\]
The chemical potential may be introduced in accordance with \cite{HK} as 
$\Omega \rightarrow e^{\frac{\mu}{T}}\Omega$, and, after the integration over 
$\psi_{x}^{\left(\pm\right)}$ fields and taking into account that 
$\lim_{a_{\tau}\rightarrow 0}(1 + a_{\tau}m)^{N_{\tau}} = e^{m/T}$,
we get for the Z(3) gauge group 
\begin{equation}
-{\cal S}_{eff}^{F} = \eta \frac{\Omega +\Omega^{*}}{2}
+ \frac{\Omega -\Omega^{*}}{2} \frac{2n_{f}}{\sqrt{3}}\phi +const  
\label{5}
\end{equation}
with $\frac{\eta}{n_f} =$
\footnote{
It seems worth noting that the fermion part of the action enjoys symmetries 
$\eta(m,\mu) = \eta(m,-\mu);~~\phi(\mu)=-\phi(-\mu)$ and $\eta(m,\mu) 
=\eta(\mu,m)$.} 
\[
\frac{2}{3} \ln \left({\rm ch}\frac{m}{T} + {\rm ch} 
\frac{\mu}{T}\right) - \frac{1}{3}\ln \left(\frac{{\rm ch}^{3}\frac{m}{T} +
{\rm ch}^{3}\frac{\mu}{T}}{{\rm ch}\frac{m}{T} + {\rm ch}\frac{\mu}{T}} -
\frac{3}{4}\right)
\]
\begin{eqnarray}
\phi(m,\mu) &=& \frac{1}{2i}\ln \left(\frac{{\rm ch}\frac{m}{T} - \frac{1}{2}
{\rm ch}\frac{\mu}{T} + i\frac{\sqrt{3}}{2}{\rm sh}\frac{\mu}{T}}{{\rm ch} 
\frac{m}{T} - \frac{1}{2}{\rm ch}\frac{\mu}{T} - i\frac{\sqrt{3}}{2}{\rm sh} 
\frac{\mu}{T}}\right); \nonumber \\
&& -\pi < \phi <\pi. \nonumber
\end{eqnarray}

In the asymptotic area $m>>T;~\mu >>T$ the real part of fermion
contribution remains essential only when $m \simeq \mu$ but outside this
region $\eta$ rapidly disappears. It is also easy to show that the
imaginary part of fermion contribution disappears both for 
$\frac{\mu}{T}\rightarrow 0$ and (if $\frac{m}{T}>>1$) for 
$\frac{\mu}{m}\rightarrow 0$
and 
\begin{equation}
\phi =\left\{ 
\begin{array}{c}
\pm\frac{2\pi}{3};~~~ \frac{\mu}{T}>>\frac{m}{T}>>1;~~~ 
\frac{\mu}{T}\rightarrow \pm\infty, \\ 
\pm\frac{\pi}{3};~~~ \frac{\left| \mu \right|}{T} = \frac{m}{T};~~~ 
\frac{\mu}{T}\rightarrow \pm\infty.
\end{array}
\right.
\end{equation}

The effective action ${\cal S}_{eff}={\cal S}_{eff}^{G}+{\cal S}_{eff}^{F}$
corresponds to the extended Potts (N=3) or Ising (N=2) model which has been 
studied in detail in \cite{DD}. Our consideration essentially differs from 
that of \cite{DD} by the specific form of 'external sources' $\eta$ and 
$\phi$. In particular, the fermion part of the action depends on $m,\mu,T$ 
only through the combinations of $\frac{m}{T}$ and $\frac{\mu}{T}$. 
In addition, in our approach the parameter $m$ may be arbitrary small.
All parameters are expressed in physical units, so the fermion part of 
the effective action does not change in 
$a_{\tau}\rightarrow 0;~~ N_{\tau}\rightarrow \infty$ limit.

\section{Mean spin approximation}

Here we apply the mean spin approach \cite{ABPZ} to compute analytically the
average value of Polyakov loop $\langle \Omega \rangle$, which
allows us to estimate $\langle \bar{\psi}\psi \rangle$ at finite
density. Let us introduce 
$\bar{\Omega}= r e^{i\theta} = \frac{1}{V}\sum_{x}^{V}\Omega_{x}$
(mean spin), where $V$ is the volume in lattice units. Adjusting the 
definition of quasiavarages \cite{NNB} to the considered case 
\begin{eqnarray}
\prec q\succ _{\bar{\Omega}} &\equiv & \sum_{(\sigma)} q 
\delta \left( 2Vr\cos \theta -2\sum_{x}
{\rm Re}\Omega _{x}\right) \nonumber \\
&& \delta \left(\frac{2\left(V r\sin\theta -\sum_{x}{\rm Im}
\Omega _{x}\right)}{\sqrt{3}}\right),
\label{6}
\end{eqnarray}
we may write for the gluon part of the effective action 
\begin{equation}
\exp \left\{-{\cal S}_{eff}^{G} \right\} \equiv \frac{\prec \exp 
\left\{ -{\cal S}^{G} \right\} \succ_{\bar{\Omega}}}
{\prec 1\succ_{\bar{\Omega }}}
\label{7}
\end{equation}
with 
\begin{eqnarray}
-\frac{1}{V}\ln \prec 1\succ _{\bar{\Omega}} &\equiv &
{\cal L} \simeq \sum_{k=0}^{2}l_{k}\ln l_{k}; \nonumber \\
l_{k} &=& \frac{1+2r\cos \left( \theta +\frac{2\pi k}{3}\right)}{3}.
\label{8}
\end{eqnarray}
The gluon part of the effective action up to $\gamma^3$ terms 
can be then written as
\footnote{
It is easy to show that the 'mean spin' method in the lowest order in $\gamma$
coincides with the simple version of the 'mean field' method.} 
\[
-\frac{{\cal S}_{eff}^{G}}{V} = \gamma r^{2} + \frac{5}{3} \gamma^{2}r^{2}
\left(1+2r\cos 3\theta -2r^{2}\right).
\]
with 
$\gamma =3\frac{1-e^{-3\tilde{\kappa}_{\tau}}}{1+2e^{-3\tilde{\kappa}_{\tau}}}$. 
Therefore we may finally write
\[
-\frac{S_{eff}}{V}=\Phi =\eta r\cos \theta +\frac{2in_{f}\phi r}{\sqrt{3}}
\sin \theta -{\cal L}-\frac{S_{eff}^{G}}{V}.  
\]

To calculate the partition function we shall find the saddle point 
$\bar{\Omega}_{0}=r_{0}e^{i\theta_{0}}$ where both 
${\rm Re}{\cal S}_{eff}(\theta,r_{0}) = \min {\rm Re}{\cal S}_{eff}$ 
and $\left| {\rm Im}{\cal S}_{eff}\left( \theta,r_{0}\right) \right|$ =
$\min \left| {\rm Im}{\cal S}_{eff}\right| $. Although at any 
$\theta \neq \pi n$ ${\rm Im}{\cal S}_{eff}\left(\theta,r_{0}\right) \neq 0$, 
it oscillates in thermodynamical limit with a very high frequency 
$\sim V\sin \theta$ and $\int_{V-\pi}^{V+\pi}{\rm Im}
{\cal S}_{eff}\left(\theta,r_{0},V^{\prime}\right) dV^{\prime}$ becomes
negligible at any large $V$. On the other hand, 
${\rm Re}{\cal S}_{eff} \left(\theta,r_{0}\right)$ gains the minimum 
at $\eta \cos\theta =\left| \eta \right| $, so, taking into account that 
$\eta \geq 0$, we may put $\theta =0$. Thereby we may conclude that the
free energy $F(\mu) \equiv -\frac{1}{T}\ln {\cal Z}$ does not depend on $\phi$ 
and differs from $F(0)$ simply by 'renormalization' 
$\eta(0)\rightarrow \eta(\mu)$. Moreover, the independence 
$F(\mu)$ of $\phi$ signals about an implicit charge symmetry 
($\Omega_{x}\leftrightarrow \Omega_{x}^{*}$ or 
$\theta \leftrightarrow -\theta$) in the model which is formally broken in 
$\Phi$.
This, in particular, leads to 
$\left\langle{\rm Im}\Omega \right\rangle \sim \frac{\partial F(\mu)}
{\partial \phi}=0$.

Therefore, one may write for the free energy 
$F\simeq -VT\max \left\{ \Phi(0,r_{0}), \Phi(0,0) \right\}$, where 
the saddle point $r_{0}$ is defined by 
$\frac{\partial \Phi \left(0,r_{0}\right)}{\partial r_{0}}=0$. Failing 
to solve such an equation precisely, we studied it in the
areas $r<<1$ and $1-r<<1$ . We find that the border line between 'ordered' 
($r_{0}=\left\langle \left| \Omega \right| \right\rangle \neq 0$) and
'disordered'($\left\langle \left| \Omega \right| \right\rangle =0$) phases
can be roughly presented
\footnote{
In fact, the border line with disordered phase has a quite complicated
structure, however, further details of its dependance on $\gamma$ and 
$\eta$ may be considered only in a more accurate approximation.} as 
$\frac{1}{2}\gamma +\eta \simeq 1$.
\begin{eqnarray}
\langle\bar{\psi}\psi\rangle &=&
\langle\bar{\psi}^{(+)}\psi^{(+)}\rangle +
\langle\bar{\psi}^{(-)}\psi^{(-)}\rangle =
\frac{T}{V_\sigma}\frac{\partial\ln{\cal Z}}{\partial m} \nonumber \\
&=& 2 \frac{1-2e^{-\frac{m}{T}}(1 - e^{-\frac{m}{T}}){\rm ch}\frac{\mu}{T}\cdot
{\rm Re}\langle\Omega\rangle}{1+e^{-3\frac{m}{T}}}.
\nonumber
\end{eqnarray}

\section{Conclusions}

This paper considers finite temperature QCD on an anisotropic lattice 
($\xi >>1$) in the approximation $SU(N) \simeq Z(N)$.

We argue that at least in above approximation the imaginary part of
the action plays a marginal role and the free energy at finite density differs
from the one at $\mu =0$ mainly by 'renormalization' of the real part of
the action: $\eta(0) \rightarrow \eta(\mu)$.

The chiral condensate in this model does not turn to zero at any 
$\frac{m}{T}\neq 0$ which indicates chiral symmetry breakdown. 
$\left\langle \bar{\psi}\psi \right\rangle $ strongly depends on 
$\langle\Omega\rangle$ for $\frac{|\mu|-m}{T} >> 1$, so the gluon 
environment plays an essential role in this region.

\end{document}